\documentstyle[12pt]{article}

\newcommand{\be}{\begin{equation}}
\newcommand{\ee}{\end{equation}}
\newcommand{\beq}{\begin{eqnarray*}}
\newcommand{\eeq}{\end{eqnarray*}}
\newcommand{\beqn}{\begin{eqnarray}}
\newcommand{\eeqn}{\end{eqnarray}}
\newcommand{\bd}{\begin{displaymath}}
\newcommand{\ed}{\end{displaymath}}
\newcommand{\resection}[1]{\setcounter{equation}{0}\section{#1}}
\renewcommand{\theequation}{\arabic{section}.\arabic{equation}}

\newcommand\Ds{\raise.15ex\hbox{$/$}\kern-.72em\hbox{$D$}}
\newcommand\ds{\raise.15ex\hbox{$/$}\kern-.59em\hbox{$\partial$}}
\newcommand\Dst{\raise.15ex\hbox{$/$}\kern-.72em\hbox{$\tilde D$}}

\newcommand\pa{\partial}
\newcommand\cond{\langle\qb\q\rangle}
\newcommand\bl{\bigl\langle}
\newcommand\br{\bigr\rangle}
\newcommand{\q}{\hbox{q}}
\newcommand{\qb}{\overline{\q}}
\newcommand{\op}{\Box^2-m^2\Box}
\newcommand{\htil}{\tilde{h}_0}
\newcommand{\km}{k^2+m^2}

\begin{document}
\setlength{\baselineskip}{21pt}
\pagestyle{empty}
\eject
\begin{flushright}
SUNY-NTG-94-28
\end{flushright}

\vskip 2.0cm
\centerline{\bf The Invariant Fermion
Correlator in the Schwinger Model on the Torus}
\vskip 2.0 cm
\centerline{J.V. Steele, J.J.M. Verbaarschot, and I. Zahed}
\vskip .2cm
\centerline{Department of Physics}
\centerline{SUNY, Stony Brook, New York 11794}
\vskip 2cm

\centerline{\bf Abstract}
\vskip 3mm

We construct the gauge invariant fermion correlator in the Schwinger model
on the torus. At zero temperature, this correlator falls off
with a rate given by the Coulomb energy of an infinitely heavy charge.
At high temperature,  the screening mass approaches $\pi T/2$,
and this in the presence of a mass gap. The fractional
Matsubara frequency arises from the action of
a pair of induced merons at high temperature that are localized over a range
on the order of the meson Compton wavelength $1/m=\sqrt{\pi}/g$. We
discuss the quenched approximation in this model, and comment on the
possible relevance of some of these results to higher dimensions.

\vfill
\noindent
\begin{flushleft}
SUNY-NTG-94-28\\
July 1994
\end{flushleft}
\eject
\pagestyle{plain}
\resection{Introduction}

A fundamental aspect of QCD is that chiral symmetry is spontaneously broken
at low temperature and restored at high temperature. As the low temperature
phase is also confining, the low temperature excitations are hadronic.
The nature of the excitations above the phase transition is still debated.
The phase is believed to be screened, but lattice simulations indicate strong
space-like correlations at $all$ temperatures (see Ref. \cite{zahed} and
references therein).

Lattice simulations as well as analytical calculations have mainly focused
on the behavior of the hadronic correlators in the space-like (Euclidean)
domain at low and high temperature \cite{lattice,hansson}. The mesonic and
baryonic correlators were
found to asymptote $2\pi T$ and $3\pi T$ respectively, with the exception of
the pion and the sigma \cite{lattice}. Lattice simulations for the quark
correlator in the Landau gauge have also been performed \cite{boyd}.

In non-gauge theories the fermionic propagator carries interesting information
on the single as well as collective excitations of the system. In gauge
theories the situation is somewhat unclear, since the naive fermion propagator
is gauge dependent. In two dimensional QCD, 't Hooft has shown
that in the axial gauge the quark propagator is infrared sensitive and
zero in the infrared limit \cite{thooft}. Einhorn later noted that the
infrared
sensitivity was directly related to the gauge sensitivity of the quark
propagator \cite{einhorn}. This feature was also noted by
Casher, Kogut and Susskind in the context of the Schwinger model \cite{casher}.
The fermion propagator has been used to probe the high temperature phase of
QCD \cite{boyd,pisarski}.

Recently, we have analyzed the mesonic correlation functions in the Schwinger
model at finite temperature both in real and Euclidean time \cite{faz}.
All correlators
were found to fall off with the meson mass $m=g/\sqrt{\pi}$ in the spatial
direction at high temperature (for $x \gg 1/m$).
Similar correlators were also investigated
by Abada and Schrock \cite{abada} and Smilga \cite{Smilga}
with somewhat similar conclusions. In this paper, we  would like to extend
the analysis to the gauge invariant fermion correlator in Euclidean space.
By construction, this correlator is a combination of a fermion
propagator and a path ordered exponential (string)
--- thus gauge invariant and free from the
ambiguities discussed above. The gauge invariant correlator reduces to the
$ordinary$ fermion correlator for (time-like) axial gauges,
barring the difficulties associated with these gauges in perturbation theory
\cite{yamagishi}. Also, the gauge invariant  correlator offers a suitable
framework for probing both chiral symmetry
breaking and confinement (screening) at zero and finite temperature
through its short and large distance limits \cite{yamagishi}.
To illustrate these points, we will use the Schwinger model on the
torus to calculate exactly this correlator both at zero and finite
temperature.
At zero temperature, the diagonal part of the gauge invariant correlator is
saturated by the fermion condensate at short distances, and falls off
exponentially at large distances. The fall-off rate is related to the screening
length of the attached string, a direct consequence of the screening
character of the QED ground state. This aside, the singularities of the
gauge invariant correlator reflect on a mass gap in the spectrum. At high
temperature, this correlator asymptotes $\exp(-\pi Tx/2)$ (for $x \gg 1/m$)
in the spatial
direction and has a free field behavior along the temporal direction
$despite$ the fact that the spectrum exhibits a mass gap at all temperatures.
We show that the deviation from $\pi T$ in the spatial asymptote is due to
a pair of induced merons.
At high temperature, the merons are localized over a range
on the order of the meson Compton wavelength $1/m=\sqrt{\pi}/g$. We discuss
the quenched approximation and suggest that it cannot be applied to this
model. The possible relevance of some of these results to higher
dimensions is discussed in our concluding remarks.

\resection{The Invariant Correlator}
Consider the Euclidean gauge invariant fermion correlator on a strip of
temporal
length $\beta =1/T$, and spatial length $L$

\be
{\cal S}_F (x, \beta ) =
-\bl \q (0)\, e^{ig\int_0^x A_\mu d\xi_\mu}\,\qb (x)\br_{\beta}.
\label{1}
\ee
Although this correlator is gauge invariant, it depends on the choice of the
path between zero and $x$. Below,  we will always choose the shortest
path on the torus staying within the interval $[0,\beta]$
in the time direction and $[0,L]$ in the space direction. The
expectation value is over the QED measure on the torus (see Sachs and Wipf
\cite{sw}
for more details). On the torus,  the gauge field, $A$, obeys the general
(Hodge) decomposition ($V=\beta L$) \cite{sw}

\be
A_\mu(x)={2\pi\over g}(-{kx^1\over V} +{h_0\over\beta},{h_1\over
L})-\varepsilon_{\mu\nu}\partial_\nu \phi.
\label{2}
\ee
The first part refers to the instanton in two dimensions with topological
charge $k$, the second part refers to the constant modes associated with the
Polyakov line, and the final part is just the transverse gauge field related
to the electric polarization. The calculations to follow are understood to be
in covariant gauges \cite{sw}.

The correlator (\ref{1}) receives contributions
only from the $k=0,\pm 1$ sectors. In higher fluxes, the zero modes, not being
absorbed by the quark fields, will make the fermion determinant vanish.
The $k=0$ sector will contribute to the
off-diagonal part (since the Green's function anti-commutes with $\gamma_5$)
and the diagonal elements will be from the $k=\pm 1$ sectors.
Inserting the explicit  form of the gauge potential
into (\ref{1}) shows that the line integral gives a phase factor
from the harmonic part (the $h$'s) and an integral from the classical
background field depending on the flux factor $k$. The latter integral vanishes
for paths exclusively along either the time or space
direction\footnote{For other choices of paths this term  would vanish upon
the limit $L\to\infty$ anyway.}.
The $\phi$-dependent term
\be
ig\varepsilon_{\mu\nu}\int_0^x d\xi_\mu
{\pa\over\pa\xi^\nu}\phi(\xi)
\label{3}
\ee
represents the quantum fluctuations around the background field and must
be averaged in the path integral.

The spinor structure in two dimensions allows $\phi$ to be factored out of the
Dirac operator
\be
\Ds=e^{g\gamma_5\phi}\Dst e^{g\gamma_5\phi}
\label{03}
\ee
where $\Dst$ is the Dirac operator in the instanton background.
With this in mind, the $k=1$ sector only contributes to the upper left entry of
the matrix, (\ref{1}). The explicit form is

\be
{\cal S}_{F,++}(x,\beta)=
-{\int d^2h\, {\psi^\dagger}_1(x)\psi_1(0) \det'(i\Dst_1) \int D\phi
e^{-{1\over2}\int \phi(\op)\phi -2\pi/m^2V} e^{ig\int_0^x A\cdot d\xi}\over
\int d^2h \det(i\Dst_0) \int D\phi e^{-{1\over2}\int \phi(\op)\phi}}.
\label{4}
\ee
The zero modes are (for arbitrary positive flux $k$ and $p$ running from 1 to
$k$)

\be \psi_{p,k}(x) = e^{-g\phi(x)} \tilde{\psi}_{p,k}(x)
\label{zero}
\ee
\bd\tilde{\psi}_{p,k}(x) = \Bigl({2k\over\beta^2 V}\Bigr)^{1/4}
U(x) e^{-\pi\tau \htil^{\ 2}/k - 2\pi i \htil (z+h_1/k)-\pi k (x^1)^2/V}
\vartheta_3(kz+h_1-i\tau\htil|ik\tau)\ed
with
\beq
U(x)=e^{2\pi i (h_0x^0/\beta + h_1x^1/L)},&&\tau=L/\beta,
 \\
\htil=h_0 -p+{1\over2},&\mbox{and}&z={x^0+ix^1\over\beta}.\eeq
For negative flux, $-k$, the sign of $\phi$ must be changed and
$\tilde{\psi}(x)
\to\tilde{\psi}^*(-x)$.
The determinants of $\Dst$ have been evaluated in Sachs and Wipf \cite{sw}. The
phase factor from the zero modes, $U(x)$, cancels with the phase factor from
the gauge line integral. The gaussian integrals can be calculated by
completing squares (see Appendix A).

After integration over $\phi$ in both numerator and denominator,
${\cal S}_{F,++}(x,\beta)$  reduces to an integration over only
the harmonic part of the potential.

\beqn
-{\sqrt{2\tau} |\eta|^2\over\beta} e^{2g^2 K_{xx}}e^{-2\pi/m^2V-\pi(x^1)^2/V
-\pi i x^0x^1/V} e^{I_3(x,\beta)}\hfill\nonumber\\
\hfill\times \int d^2h e^{-2\pi\htil^2\tau +2\pi i\htil \overline{z}}
\vartheta_3(\overline{z}+h_1+i\tau \htil)\,\vartheta_3(h_1-i\tau\htil)
\label{5}
\eeqn
As in \cite{sw}, $\eta$ refers to Dedekind's eta-function evaluated at
$i\tau$.
The $h_1$ integration may be done after expressing the theta functions as
Fourier series. This gives a Kronecker delta that combines the two sums into
one. This one sum then extends the $h_0$ integration from the segment $[0,1]$
to $(-\infty,\infty)$ which allows a Gaussian integration to be done.
Identifying the fermion condensate \cite{sw}

\be
\bl\qb\q\br_{\beta}= -{2 |\eta|^2\over\beta} e^{2g^2K_{xx}}e^{-2\pi/m^2V},
\label{6}
\ee
the answer for the $k=1$ sector is

\be
{\cal S}_{F,++}(x,\beta)=
\bl \qb (x)\, e^{ig\int_0^x A\cdot d\xi}\,\q (0)\br_{\beta}^{k=1}=
{\bl\qb\q\br_{\beta}\over2}\,e^{I_3 ({x}, \beta )-\pi((x^0)^2 + (x^1)^2)/2V}
\label{7}
\ee
with $I_3$ given by

\be
I_3 ({x}, \beta ) =
\frac {g^2}2 \int_0^x d\xi_{\mu}d\xi'_{\mu} \Box K_{\xi\xi'}\, .
\label{8}
\ee
Since $\Box K =(\Box - m^2)^{-1} $, the integration may be carried
out by expanding in complete eigenfunctions of $\phi$ as done in Appendix B.
The result, for $L\to\infty$, in the temporal direction is ($t>0$)
\be
I_3 (t, \beta )=-{\pi mt\over4}-{m^2\over2}\int_0^\infty
{dk\over(\km)^{3\over2}}
\left( e^{-t\sqrt{\km}}-1 +2{(\cosh t\sqrt{\km}-1)\over e^{\beta\sqrt{\km}} -1}
\right)
\label{9}
\ee
and in the spatial direction

\be
I_3 (x^1, \beta )=-m^2\int_0^\infty {dk\over\sqrt{\km}} {\sin^2{kx^1
\over2}\over k^2} \coth {\beta\over2}\sqrt{\km}.
\label{10}
\ee
Note that in the spatial direction, the high temperature limit of $I_3$ is
($x^1>0$)

\bd
-{\pi x^1\over 2\beta} +{\pi\over2m\beta}(1-e^{-mx^1}).
\ed
This results in a screening mass $\pi T/2$. We want to note
that this factor is not from the Dirac string but from the $K_{x0}$ term of the
$\phi$ integration.

For the $k=0$ sector, the propagator is needed. In the upper right
corner of the matrix, it is \cite{faz,jaz}

\be
G_{+-}(0,x)=e^{g(\phi(x)-\phi(0))}\, U^\dagger (x) {i|\eta|^3\over \beta}
{\vartheta_4(z-H)\over\vartheta_1(z)\vartheta_4(H)}\, e^{2\pi i h_0z}
\label{010}
\ee
with $H=h_1-i\tau h_0$
and $G_{-+}$ is just obtained by taking the complex conjugate and the
coordinates to their negative. Writing the determinant as

\bd
\det(i\Dst_0)=\Bigl|{\vartheta_4(H)\over\eta}\Bigr|^2 e^{-2\pi\tau
h_0^{\ 2}},
\ed
the $h$ integrations may be done and the square completed in the $\phi$
integration to give
\bd
{\cal S}_{F,+-}(x,\beta)=
-{i|\eta|^3\over\beta} {e^{2g^2(K_{xx}-K_{x0})}\over\vartheta_1(z)}
e^{I_3 ({x}, \beta )-\pi z^2/2\tau}.
\ed
Using the form for the bosonic propagator

\bd
g^2 K_{xy}=\pi\Delta_m(x-y) +\ln\bigl|{\eta\over\vartheta_1(z_1-z_2)}
\bigr|^{1\over2} +{\pi\over 2V}(x^1-y^1)^2+{\pi\over m^2V}
\ed
found elsewhere \cite{faz,jaz}, the result is

\be
{\cal S}_{F,+-}(x,\beta)=
i{\bl\qb\q\br_{\beta}\over2} \frac{|\vartheta_1(z)|}{\vartheta_1(z)}
e^{I_3(x,\beta)-2\pi\Delta_m(x)} e^{-\pi((x^0)^2+(x^1)^2)/2V-\pi i x^0 x^1/V }
\label{11}.
\ee
In the thermodynamic limit this result can be simplified further. The phase of
the $\vartheta_1$ function is then given by
\be
e^{-i\varphi}\equiv\frac{|\vartheta_1(z)|}{\vartheta_1(z)}=\frac{|\sin(\pi z)|}
{\sin(\pi z)}
\ee
which is $\varepsilon(x^0)$ for purely temporal paths and $-i\varepsilon(x^1)$
for purely spatial ones. For $L\rightarrow\infty$
$\Delta_m$ can replaced by its Poisson resummation:

\bd
-2\pi \Delta_m(x,\beta) =
\sum_n K_0\left(m \sqrt{(x^0-\beta n)^2+(x^1)^2}\right).
\ed
The result for the lower left entry in the matrix is just the conjugate of
equation (\ref{11}).
Putting together (\ref{7}) and (\ref{11}) we finally have for the gauge
invariant fermion correlator in
the Schwinger model at finite temperature in the infinite
volume limit

\be
{\cal S}_F(x,\beta)=
{\bl\qb\q\br_{\beta}\over2}\,e^{I_3 ({x}, \beta )}
\left( \pmatrix{e^{i\theta}& 0\cr 0& e^{-i\theta}\cr}
+ i e^{-2\pi\Delta_m(x,\beta)}
\pmatrix{0& e^{-i\varphi}\cr e^{i\varphi}& 0\cr}\right)
\label{12}
\ee
where $\theta$, the vacuum angle, has been added for completeness.
For $x\to 0$ (\ref{12}) is consistent with the operator product expansion.
The explicit form for $I_3$ is given in (\ref{9}) and (\ref{10})
for both temporal and
spatial directions. Note that we can rewrite those expressions
in the equivalent form

\be
I_3 (t, \beta ) = -\frac{\pi m t}4 -\int {d^2p\over2\pi} \,\delta (p^2-m^2 )
\frac {\pi m^2}{p_0^{\ 2}}\left(\theta (p_0 ) + n_{\beta} \right)
\left( e^{-p_0 t} -1\right)
\label{13}
\ee
along the temporal direction, and

\be
I_3 (x^1, \beta ) = +\int \frac{d^2p}{2\pi} \,\delta (p^2-m^2 )
\frac {\pi m^2}{p_1^{\ 2}}\left(\theta (p_0 ) + n_{\beta} \right)
\left( e^{ip_1 x^1} -1\right)
\label{14}
\ee
along the spatial direction. Here
$n_{\beta }= \left( e^{\beta\sqrt{k^2+m^2}}-1\right)^{-1}$ is the Bose
distribution. The expressions (\ref{13}-\ref{14}) are amenable to a spectral
analysis. At zero temperature, (\ref{14}) can be rewritten
in the form

\be
I_3 = -\frac{\pi m x^1}4 -\int \frac{d^2p}{2\pi} \,\delta (p^2-m^2 )
\frac {\pi m^2}{p_0^{\ 2}}\,\,\theta (p_0 ) \,
\left( e^{-p_0 x^1} -1\right)
\label{15}
\ee
as expected from $O(2)$ invariance. The result for zero temperature
in the spatial direction has also been obtained in the gauge $A_0 = 0$
\cite{Becher}. Although the Schwinger factor is absent in that case it
seems to us that the present approach is more transparent.

\resection{Dimensional Reduction}

At zero temperature, we note that the trace of the gauge
invariant correlator
reduces to $\cond_{\beta}\cos\theta$, which is the  expected fermion
condensate at finite temperature and nonzero vacuum angle.
At large Euclidean separations, it falls off as $e^{-m\pi |x|/4}$ in both
the spatial and temporal directions. We note that

\be
\frac {\pi m}4 = \frac {g^2}2 \int \frac {dp}{2\pi} \frac 1{p^2+m^2}
\label{16}
\ee
is just the Coulomb energy of an infinitely heavy charge. The Coulomb
energy follows from the screening of the line integral present in the gauge
invariant correlator in the QED vacuum. It is a purely ``kinematical" term.
This aside, the expressions (\ref{13}) and (\ref{15}) show that the
singularities of the
gauge invariant correlator are related to the mass gap $m$ with a form factor
${\pi m^2}/{p_0^{\ 2}}$.
The off-diagonal part of the gauge invariant correlator reduces
to the free fermion propagator at short distances $i\gamma\cdot x/|x|^2$.
The large distance behavior is similar to the diagonal part. Our result
(\ref{12}) at
zero temperature is in disagreement with the brief analysis by Sachs and Wipf
of a related correlator (\cite{sw}, section 6).

At finite temperature, we note that the diagonal part of the gauge invariant
correlator reduces to the temperature dependent condensate at zero separation,
while the off-diagonal part gives the usual free fermion propagator. At
high temperature, the gauge invariant correlator along the temporal direction
exhibits a free field behavior. Along the spatial directions,
the gauge invariant correlator falls off at a rate which is given by
$e^{-\pi T |x|/2}$ despite the fact that the spectrum has a mass gap
$m=g/\sqrt{\pi}$.

To understand this behavior,
we note that at high temperature the model
dimensionally reduces to one dimension since ($\omega_n =2\pi n T$).

\be
\phi (t, x) =\sum_n e^{-i\omega_n t}\,\phi_n (x) \sim \phi_0 (x)
\label{17}
\ee
Therefore, $A_1\sim 0$ and

\be
A_0 (x)\sim \frac {2\pi h_0}{g\beta} -\partial_1 \phi_0 (x).
\label{18}
\ee
With this in mind, the diagonal part of the gauge
invariant correlator reduces to

\be
\bl \qb (0,x)\, e^{ig\int_0^x dx' A_1 (0,x') }\,\q (0,0)\br_{\beta} \sim
\bl \qb (0,x)\,\q (0,0)\br_{\beta}.
\label{19}
\ee
Because of the trace over Dirac indices, only the zero modes contribute to
(\ref{19}). In the dimensionally reduced theory, the expectation value becomes

\be
\bl \qb (0,x)\,\q (0,0)\br_{\beta} \sim
\bl e^{-g (\phi_0 (x) +\phi_0 (0) )}\br_{\beta}.
\label{20}
\ee
The expectation value is with respect to the dimensionally reduced action

\be
S_{0+1} = \beta\int dx \frac 12 \phi\left( \nabla^4 -m^2\nabla^2 \right) \phi.
\label{21}
\ee
The integral being quadratic in $\phi$ can be performed at once using the
saddle point equation.

\be
\left(\nabla^4 -m^2 \nabla^2\right) \phi_{0,*} (z) =
-gT\left(\delta (z-x) +\delta (z-0)\right)
\label{22}
\ee
The solution is an induced instanton (first introduced in the context
of mesonic correlation functions \cite{Nielsen}).
\be
\phi_{0,*} (z) = \frac{\pi T}{2g}
\left(\frac 1m e^{-m |z|} + |z| +\frac 1m e^{-m|z-x|} +|z-x|\right)
\label{23}
\ee
Inserting (\ref{23}) into (\ref{20}), (\ref{21}), and using (\ref{22})
yields ($|x|\rightarrow\infty$)

\be
\bl \qb (0,x)\,\q (0,0)\br_{\beta} \sim
e^{-\frac{\pi T}2 \left( |x| +\frac 1m +\frac 1m e^{-m|x|}\right)}\sim e^{-\pi
T |x|/2 -\pi T/m}.
\label{24}
\ee
For the off-diagonal part the induced instanton
gives rise to a factor $\exp(\pi Tx/2)$ which together with a factor
$\exp(-\pi T x)$ from the free propagator results in the same asymptotics.
As final answer we find

\be
\bl \q (0,0)\,\qb (0,x)\br_{\beta} \sim-
e^{-\pi T |x|/2-\pi T/m}\left(\frac 12 {\bf 1}
-\frac{i{\bf \gamma}^1}{2\beta}\right)
\label{25}
\ee
at high temperature.

We remark that for $x \gg 1/m$ the induced instanton (\ref{23})
is a linear superposition of
two merons. Indeed, the classical field

\be
\phi_{0,*} (z) = \frac{\pi T}{2g}
\left(\frac 1m e^{-m |z|} + |z|\right)
\label{26}
\ee
gives rise to the following potential ($h_0 =-1/4g$)

\be
A_0 (z) = \frac{\pi T}{2g}{\rm sgn} (z) \left(e^{-m |z|} -1 \right)
-\frac {\pi T}{2g}
\label{27}
\ee
with $A_0(-\infty ) = 0$ and $A_0 (+\infty ) =-\pi T/g$. The topological charge
carried by (\ref{26}) is

\be
\nu = \frac {g\beta}{2\pi }\int dx \, E(x) = -\frac{g}{2\pi T}
\left( A_0(+\infty ) -A_0 (-\infty )\right) = \frac 12
\label{28}
\ee
one-half of the instanton charge.
In the Schwinger model the merons are induced
(refer to (\ref{22})) and localized over the meson Compton wavelength $1/m$
since the electric field $E (x)\sim Te^{-m |x|}$.
When $x$ becomes on the order of $1/m$, the two merons merge into a single
instanton. The solution for $x=0$ was studied in \cite{Smilga}; its classical
action is responsible for the temperature dependence of the condensate.

Mesonic correlation functions can also be analyzed along these lines
\cite{Smilga}. In that case the action of the classical solution cancels
the $\exp(-2\pi Tx)$ asymptotics of the free propagator resulting
in an exponential fall off determined by the meson mass. In the present case
both the classical solution and its action are a factor two smaller (as
compared to the mesonic action in the $k = 2$ sector)  and in the $k =\pm 1$
sectors there is no additional $\exp(-\pi Tx)$ dependence
resulting in the asymptotic screening mass $\pi T/2$.
In the $k=0$ sector
the action of the two meron configuration gives rise to a factor
$\exp(\pi Tx/2)$ but in this case a factor $\exp(-\pi Tx)$ comes from
the free propagator. Merons at high temperature therefore
give rise to fractional Matsubara frequencies in the Schwinger model.

\resection{The Quenched Approximation}

The quenched approximation for the Schwinger model has been discussed
extensively in the literature \cite{quenched} and a number of contradictory
statements have been made. Part of the confusion will be discussed below.
Before this, however,
we would like to stress the fact that in the quenched approximation and at zero
temperature there are no gauge degrees of freedom. The fermionic ground state
is trivial, and as such $\cond$ should be zero. In the absence of the Dirac
sea, there is no (axial)
anomaly in the quenched approximation in any dimension. At finite temperature,
and in the Euclidean formulation, the Polyakov lines play the role of non-local
gauge degrees of freedom. Dimensional reduction arguments suggest that the
high temperature phase in the quenched approximation is still trivial.

Having said this, we may $formally$ calculate $\cond$ for
fixed fermionic mass $\mu$, number of flavors $N_F$ and two volume $V$, and
consider the limits
$\mu \rightarrow 0$, $N_f \rightarrow 0$ and $V\rightarrow \infty$.
Some of these calculations have already appeared in the literature, with
some confusion related to the order of the limits.
In general, each pair of these limits does not commute. Let us first discuss
the mass ($\mu$) dependence. The important parameter is
$\xi \equiv \mu V \cond$. For $\xi\ll 1$, the mass is
much
smaller than the smallest eigenvalue and, as a consequence, only the sectors
with topological charge equal to $\pm 1$ contribute to the condensate. For
$\xi \gg 1$ the mass is much larger than the smallest eigenvalue, and the
condensate receives contributions from many different topological sectors
\cite{LS}.
Remarkably, in both cases the value of the condensate is the same.
To be more precise, if $\xi \ll 1$, the primed sum in the expression
for the condensate
\be
\cond = \frac 1V\left < \left ( \frac 1{i\mu}+{\sum}'
\frac 1{\lambda_n + i\mu} \right ) i\mu \prod_n (\lambda_n^2 + \mu^2) \right >
\label{condensate}
\ee
is of order $\mu$ and vanishes with respect to the contribution from the
zero eigenvalue. For $\xi \gg 1$ the second term becomes dominant.

In the quenched approximation, the gauge invariant correlator receives
contributions from all $k$-sectors since the restriction from the zero modes in
the integral is no longer present. The same effect occurs by taking a
sufficiently large quark mass, and we expect that in this limit the
quenched approximation is valid. In the opposite case, $\xi \ll 1$, the
condensate diverges in the chiral limit. This is clear from (\ref{condensate})
because the factor $i\mu$ from the determinant is absent in this case.
The extra weight in the bosonic integration
${1\over2}m^2\,\phi\Box\phi$, which originally was from the fermionic
determinant \cite{sw} is no longer there. This results in a different
bosonic Green's function. With this in mind, the fermionic propagator reads
\be \q(0) \qb(x)=\sum_n{\psi_n(0)\overline{\psi}_n(x)\over \lambda_n +i\mu}
\label{29}
\ee
with $\mu$ being a regulator for the zero modes which will be taken to zero
after $L\to\infty$. In the Schwinger model and also for QCD with one flavor
the result for the condensate is independent of the order
of the limits $\mu\rightarrow 0$  and $V\rightarrow \infty$. Assuming that
this is also the case for the Schwinger model in the quenched approximation,
we choose $\mu \ll 1/ \sqrt V$ so that only the $k=\pm 1$ sectors contribute
to the condensate in the quenched approximation. Therefore, for $\xi \ll 1$,
the fermion condensate becomes
\be
\cond_Q = \frac{2i}{\mu V}\frac{\sum_{k>0} k
e^{-2\pi k^2/m^2V} } {\sum_{k>0} e^{-2\pi k^2/m^2V}}.
\ee
In the limit $V\rightarrow\infty$ this sum can be calculated:
\be
\cond_Q = \frac {i \sqrt 2}{\pi} \frac m{\mu \sqrt V}
\ee
in agreement with an analysis using the proper time representation of the
fermion propagator \cite{Guerin-Fried}. The fact that the condensate
diverges for $\mu \rightarrow 0$ seems to be in contradiction with arguments
presented at the beginning of this section. However, the condensate as
calculated in (4.3) follows from the trace over $all$ fermionic states
and cannot be identified with the condensate that would be
obtained in the absence of the Dirac sea (which is zero).

The gauge invariant fermion
correlator cannot  be calculated exactly in the quenched approximation.
The Wilson line --- being different in the quenched approximation ---
shows an area law (quenched) as opposed to a perimeter law (unquenched).
As such, it would be interesting to evaluate the ratio of the invariant
correlator to  the fermion condensate in the quenched approximation.

\resection{Conclusion}

We have explicitly constructed the gauge invariant fermion correlator for the
Schwinger model on the torus. We have shown that at zero temperature, the
result is $O(2)$ invariant and asymptotes $e^{-\sigma_P |x|}$ at large
separations with $\sigma_P =\pi m/4$ being the screening mass associated with
the
line integral. This screening mass is purely kinematical, and relates directly
to the perimeter law of the Wilson loop at zero temperature

\be
\bl e^{ig\oint_C A\cdot d\xi}\br_{\beta} \sim  e^{-\sigma_P \,\,P}
\label{34}
\ee
where $P$ is the perimeter spanned by the $large$ loop $C$. At short distances,
the diagonal part of the gauge invariant correlator reduces to the fermion
condensate, while the off-diagonal part reduces to the free fermion propagator.
After proper subtractions, the gauge
invariant correlator displays singularities that
are related to the mass gap in the theory. These singularities may
be related to heavy-light systems, with the string playing the role of an
infinitely heavy charge. At finite temperature, we have shown that the diagonal
part of the gauge invariant correlator at short distances reduces to the
fermion condensate and asymptotes $\exp(-\pi Tx/2)$
at large spatial separations, even though
the spectrum exhibits a mass gap at all temperatures. The
fractional Matsubara frequency results from the action of localized
merons at high temperature. We have shown that the quenched approximation
does not work in the Schwinger model. These results suggest that this is also
the case for QCD with massless quarks. However, it may very well be that
for nonzero quark masses the effect of the fermion determinant can be
ignored for most observables. The relation of
these results to the gauge invariant correlator in real-time will be
discussed elsewhere \cite{jz}.
We briefly note that the gauge
invariant correlator is not a periodic function in the temporal direction.

Are the above results relevant for four dimensional QCD? We do not know.
We suspect however, that the gauge invariant correlator for
QCD with one flavor should reflect on the
fermion condensate along the diagonal. It should asymptotically approach
$e^{-\sigma_P |x|}$
at zero temperature in all directions with $\sigma_P$ as the perimeter-law
coefficient in the Wilson loop. It would be interesting to check this
point by lattice QCD simulations and relate the results
to the (renormalized) Coulomb energy in four dimensions. This kinematical
term aside, we suspect that the rest of the correlator should fall off at
a rate determined by the heavy-light bound states of QCD. At finite
temperature, the diagonal part should be proportional to the finite
temperature fermion condensate. In particular, it should vanish in the chirally
symmetric phase. Lattice simulations of the mesonic screening masses together
with dimensional reduction arguments \cite{zahed,hansson}
suggest that at high temperature the
screening masses asymptote $2\pi T$. Repeating the dimensional reduction
arguments, and barring induced merons at high temperature, we would conclude
that the invariant fermion correlator asymptotes $\pi T$ and not a fraction of
$\pi T$ in QCD. We recall, however, that merons in QCD have been proposed
as candidates for understanding confinement at zero temperature
\cite{gross}. Their
introduction at zero temperature is, however, ad hoc, as they carry infinite
action before smearing. It would be interesting to see whether they could be
induced  and localized (have a finite action) in QCD at high temperature.
Lattice simulations in these directions could be helpful.

\vglue 0.6cm
{\bf \noindent  Acknowledgements \hfil}
\vglue 0.4cm
 The reported work was partially supported by the US DOE grant
DE-FG-88ER40388.

\vfill

\newpage
\noindent
\renewcommand{\theequation}{A.\arabic{equation}}
\setcounter{equation}{0}
{\Large\bf Appendix A}

We detail in this appendix the calculations that led to (\ref{5})
in the text. The factors associated with the exact Dirac operator in
(\ref{03}), along with the line integral (\ref{3}), act as effective
sources in the action for the $\phi$ field.

\beqn
S_{\mbox{eff}}[\phi] &=&
\int d^2y \bigl\{ {1\over2} \phi(\op)\phi +\phi J\bigr\} \nonumber\\
&=& {1\over2} \int d^2y
\{\tilde{\phi}(\op)\tilde{\phi}-J(\op)^{-1} J\},
\label{A1}
\eeqn
with
\bd
{1\over g} J(y)\equiv e_x \delta^2(y-x) +e_0\delta^2 (y)
+i\varepsilon_{\mu\nu}\int_0^x d\xi_\mu{\pa\over\pa\xi^\nu}\delta^2(y-\xi).
\ed
Here, the $e_i$'s are either plus or minus depending on in which $k$-sector the
correlator is evaluated. They have the same sign in the $k =\pm 1$ sectors
and opposite signs for the propagator in the $k = 0$ sector (\ref{010}).
The fermions contribute strongly at the endpoints
(as delta functions) but the gauge field's contribution is smeared out along
the transverse direction of the path.

The source-source term in equation (\ref{A1}) given in terms of the bosonic
Green's function, $K =(\op)^{-1}$, and including the extra minus
sign in the exponent is

\be g^2(K_{xx} + e_x e_0 K_{x0}) + i e_x I^{x}_1 + i e_0 I^{0}_1 +
I_2\label{A2}
\ee
with

\bd I_1^{x_i}=-g^2 \, \varepsilon_{\mu\nu}\int_0^x d\xi_\mu\pa_{\xi^\nu} \,
K_{\xi x_i}
\ed
\bd I_2={g^2\over2}\varepsilon_{\mu\nu}\varepsilon_{\rho\sigma}\int_0^x\int_0^x
d\xi_\mu d\eta_\rho \pa_{\xi^\nu}\pa_{\xi^\sigma} \, K_{\xi\eta}.
\ed
The first term in equation
(\ref{A2}) is just the self-energy contribution from the
fermions along the path.
The vanishing of $I_1$ follows from the mode expansion of the Green's
function. Due to the identity
$\varepsilon_{\mu\nu}\varepsilon_{\rho\sigma}=
\delta_{\mu\rho}\delta_{\nu\sigma}-\delta_{\mu\sigma}\delta_{\nu\rho}$,
the self-energy of the gauge field, $I_2$, may be rewritten as follows

\bd
I_2= g^2\bigl( K_{xx}-K_{x0} \bigr) +\frac {g^2}2
\int_0^x d\xi_\mu d\xi'_\mu \Box
K_{\xi\xi'} \equiv g^2\bigl( K_{xx}-K_{x0}\bigr) +I_3 ({x} ,\beta ).
\ed
giving finally in our case for the bosonic contribution in the exponent

\bd
2g^2(K_{xx}+ K_{x0}) + I_3(x,\beta)
\ed
in the $k=\pm 1$ sectors and
\bd
2g^2(K_{xx}- K_{x0}) + I_3(x,\beta)
\ed
in the $k = 0$ sector.

\bigskip
\noindent
\renewcommand{\theequation}{B.\arabic{equation}}
\setcounter{equation}{0}
{\Large \bf Appendix B}
\bigskip

In this appendix, we detail the calculation that leads to the explicit form of
the $I_3 ({x}, \beta )$ integral both along the temporal and spatial
directions.
The double integral of the bosonic Green's function needed in Appendix A is

\bd
I_3 (t, \beta )={g^2\over2}\int_0^t dt'\, dt'' \Box K(t'-t'',x)|_{x=0}\ed
in the temporal direction. Since $K=(\op)^{-1}$, an additional $\Box$ in the
numerator leaves only the massive part of the propagator.
Using a complete set of states of $\phi$ this may be rewritten as
\beq I_3 (t, \beta )&=&
-{\pi m^2\over 2V}\int_0^t\int_0^t dt'\, dt''
\sum_{(n_0,n_1)\ne (0,0)} {e^{2\pi i
n_0(t'-t'')/\beta}
\over \left(2\pi n_0/\beta\right)^2 + \left(2\pi n_1/L\right)^2 + m^2}\\
&=& -{\pi m^2 t^2\over 2V}\sum_{n\ne 0} {1\over \left(2\pi n/L\right)^2 +
m^2}\\
&&-{\pi\over2\tau}\bigl({m\beta\over2}\bigr)^2\sum_{n_1=-\infty}^{\infty}
\sum_{n_0\ne0} {\sin^2{\pi nt/\beta}\over (\pi n_0)^2\left((\pi n_0)^2 + (\pi
n_1/\tau)^2 + (m\beta/2)^2\right)}\eeq
The $n_0$ sum in the second term may be done by integrating both sides of
\bd
\sum_{n\ne0} {\cos 2\pi n\phi\over (\pi n)^2 + a^2}={1\over a} {\cosh
a(1-2|\phi|)\over\sinh a}-{1\over a^2}
\ed
with respect to $\phi$ twice. The result in the temporal direction for
finite $L$ is
\be
{\pi t^2\over 2V}-{\pi m|t|\over4}\coth {mL\over2}-{\pi m^2\over2L}
\sum_{n=-\infty}^{\infty} {\coth{\beta\over2}\xi(n) \left(\cosh t\xi(n)-
1\right)-\sinh |t|\xi(n)\over \xi^3(n)} \label{temp}\ee
\bd\mbox{with}\qquad \xi(n)=\sqrt{(2\pi n/L)^2+m^2}.
\ed
Taking the limit $L\to\infty$, the sum turns into an integral and gives
\bd
I_3 (t, \beta )
=-{\pi m|t|\over4}-{m^2\over2}\int_0^\infty {dk\over(\km)^{3\over2}}
\Bigl[\coth{\beta\over2}\sqrt{\km}
(\cosh t\sqrt{\km}-1)\ed
\bd\hfill -\sinh |t|\sqrt{\km}\Bigr].
\ed
The part in the brackets may be rewritten as
\bd e^{-|t|\sqrt{\km}}-1+{2\over e^{\beta\sqrt{\km}} -1}(\cosh
t\sqrt{\km}-1)\ed
showing the Bose-Einstein number occupation factor for finite temperature with
a form factor.

For the spatial result, all that is needed is to exchange $x$ with $t$ and $L$
with $\beta$ in equation \ref{temp}. Taking $L\to\infty$ in that result
produces
\beq
I_3 (x^1, \beta ) &=& -{\pi m|x^1|\over4}\coth {m\beta\over2}
+{\pi m^2\over2\beta}
\sum_{n=-\infty}^{\infty} {1-e^{-|x^1|\sqrt{(2\pi nT)^2+m^2}}\over \bigl((2\pi
nT)^2+m^2\bigr)^{3\over2}}\\
&=& -m^2\int_0^\infty {dk\over\sqrt{\km}} {\sin^2{kx^1\over2}\over k^2} \coth
{\beta\over2}\sqrt{\km}.\eeq
In the first expression $T=1/\beta$ was used to show the Matsubara form. The
second expression may be obtained by expanding the hyperbolic cotangent in the
first expression as a series, combining the two series, and noting that
\bd \int_0^\infty {dk\,\sin^2 {ka\over2}\over k^2 (k^2+b^2)}={\pi a\over b^2} -
{\pi\over4b^2}(1-e^{-ab})\ed
to produce the integral.
\newpage

\eject
\setlength{\baselineskip}{15pt}

\end{document}